\documentclass[emulateapj]{emulateapj}
\begin{document}
\shorttitle{Terrestrial Zone Disks in h and $\chi$ Persei }
\shortauthors{Currie, T. et al.}
\title{Terrestrial Zone Debris Disk Candidates in h and $\chi$ Persei }
%\title{Possible Evidence for Terrestrial Planet Formation from Strong IRAC-excesses in h \& $\chi$ Persei }
%\title{Strong IRAC-excess sources in h \& $\chi$ Persei: Possible Evidence for Terrestrial Planet Formation}
\author{Thayne Currie\altaffilmark{1}, Scott J. Kenyon\altaffilmark{1}, George Rieke\altaffilmark{2}, 
Zoltan Balog\altaffilmark{2, 4},\& Benjamin C. Bromley\altaffilmark{3}} 
\altaffiltext{1}{Harvard-Smithsonian Center for Astrophysics, 60 Garden St. Cambridge, MA 02140}
\altaffiltext{2}{Steward Observatory, University of Arizona,  933 N. Cherry Av. Tucson, AZ 85721}
\altaffiltext{3}{Department of Physics, University of Utah, 201 JFB, Salt Lake City, UT 84112}
%\altaffiltext{4}{Department of Physics \& Astronomy, University of California-Los Angeles, Los Angeles, CA 90095}
\altaffiltext{4}{on leave from Dept. of Optics and Quantum Electronics, University of Szeged, H-6720, Szeged, Hungary}
\email{tcurrie@cfa.harvard.edu}
\begin{abstract}
We analyze 8 sources with strong mid-infrared excesses in the 13 Myr-old double cluster 
h and $\chi$ Persei.  New optical spectra and broadband SEDs (0.36-8 $\mu m$) are 
consistent with cluster membership.  We show that material with T$\sim$300-400 K and L$_{d}$/L$_{\star}$
$\sim$10$^{-4}$-10$^{-3}$ produces the excesses in these sources.  Optically-thick blackbody disk models - including 
those with large inner holes - do not match the observed SEDs.  The SEDs of optically-thin 
debris disks produced from terrestrial planet formation calculations match the observations 
well.  Thus, some h and $\chi$ Persei stars may have debris from terrestrial zone planet formation.
\end{abstract}
\section{Introduction}
Radiometric dating (Yin et al. 2002) suggests 
that the Earth had $\sim 90\%$ of its mass $\sim$ 30-40 Myr after the formation of the Sun.
  Terrestrial planet formation models suggest that 
most of this mass accumulation was complete by $\sim$ 10 Myr and was 
dominated by mergers of moon-mass oligarchs (Kenyon \& Bromley 2006, 
hereafter KB06; Chambers 2001).  How terrestrial planets form around 
other stars is unresolved. 

Optically-thin infrared (IR) emission from collisionally-produced debris 
is an important observational signature of planet formation (Kenyon \& Bromley 
2004, hereafter KB04; Rieke et al. 2005, hereafter R05; KB06).  
For a 1-2 M$_{\odot}$ star, terrestrial planet formation should
produce debris (and hence strong 5-10 $\mu m$ emission) for $\approx$ 
10-20 Myr.  The 5-10 $\mu m$ signature of debris 
production should then begin to fade (KB04; R05; KB06).
While many debris disk surveys (R05; Gorlova et al. 2006; Su et al. 2006) have discovered evidence for 
Kuiper-belt analogues from 24--70 $\mu m$ excess emission around 
early A stars, there have been fewer searches for terrestrial zone debris disks around 
$\sim$ 1-2 $M_{\odot}$ stars.
Previous searches (e.g. Meyer et al. 2006; Silverstone et al. 2006)
have also concentrated on protoplanetary disks
$\lesssim$ 5 Myr old, or on the evolution
of classical debris disks (ages $\gtrsim$ 30 Myr) when 
planet formation is nearly complete and debris production 
is declining.

The double cluster h and $\chi$ Persei (d=2.34 kpc, $\sim$ 13 Myr old) provides an ideal probe of 
terrestrial planet formation.  The clusters contain $\gtrsim$ 5000 stars with 
masses $\gtrsim$ 1.3 $M_{\odot}$, an order of magnitude larger population than other
 nearby, evolved clusters.
The recent IRAC 3.6-8 $\mu m$ survey of h and $\chi$ Persei by Currie et al. 
(2007; hereafter C07) discovered a large population of sources 
with IR excess emission in the IRAC bands (up to 4-8\% at 8$\mu m$).  C07 showed 
 that the frequency of disk emission is larger for less massive stars
and at greater distances from the central star. 
The IR excesses are bluer, $K_{s}$-[IRAC]$\sim$ 0.5-1.5, than those
for Class II T Tauri stars (e.g. Kenyon \& Hartmann 1987; 
hereafter KH87), which suggests that the disks are less optically thick and 
perhaps in an evolved state.

The goal of this paper is to constrain the source of the excess IR emission  
in h and $\chi$ Per stars by modeling the SEDs and IRAC/MIPS colors of 
sources with high quality IR/optical photometry and spectra.  
In \S 2 we describe our sample and explain our 
SED fitting.  In \S 3 we show that the IR excess is best
explained by debris from terrestrial planet 
formation. Both photospheric emission and optically-thick disk models 
fail to explain the SED slopes and IRAC colors.  Optically-thick disk + 
inner hole models are marginally consistent with photometry $\le$ 5.8 $\mu m$ but are not 
consistent with either the 8$\mu m$ ([8]) data or MIPS data.
% and search for debris disks 
%around lower-mass stars.  
%If confirmed, these results provide evidence for 
%terrestrial planet formation around other stars.
%\section{Source Selection, Dereddening, and Spectral Typing}
\section{Infrared Photometry and Spectral Typing}
The C07 survey covered $\sim$ 0.75 sq. deg. centered on h and 
$\chi$ Persei.  We select stars with J$\le$16
 and with 'strong excesses' at multiple IRAC bands, $K_{s}$-[5.8] $\ge$ 0.5 
(or $K_{s}$-[4.5] $\ge$ 0.5 in absence of [5.8] data) and $K_{s}$-[8]$\ge$0.75. 
All of these sources have K$_{s}$$\ge$13.5.  In this sample, 1,915 have 
5$\sigma$ detections at [5.8] (or [4.5]) and [8], lie along the 
h and $\chi$ Per isochrone, and were not flagged for contamination 
by background galaxies (see \S 2.3 of C07 for criteria).  
Eight of these sources (see Table 1) have 'strong excesses' and  
have high-quality optical photometry from Keller et al. (2001; UBVI) 
or Slesnick et al. (2002; V band only).  

We derive a worst-case probability that the [5.8] 
and [8] emission from these 8 sources are contaminated by z=0.1-0.4 
PAH-emission galaxies that escaped flagging and could masquerade as a 'strong excess' source.  
We assume that all sources from our population 
are as faint as our faintest source ($K_{s}$=15) and that any superimposed galaxy with [5.8] $\le$ 15 
 will trigger selection of one of the 1,915 sources. 
We also assume that all galaxies are PAH-emission galaxies, that none were removed by flagging, and
that each star covers an 'area' of $\pi$2$^{2}$ square arcseconds. 
Using the galaxy number counts for the Bootes field from Fazio et al. 
(2004; $\sim$ 1,165/sq. deg. for [5.8]$\le$15), the expected
number of contaminated sources is N$_{p}$$\sim$ 2.2 ($\pi$2$^{2}$$\times$1165/(3600$^{2}$)$\times$1915).
If we adopt the number density of PAH-emission galaxies for this field 
(Stern et al. 2005) instead of the number density of \textit{all}
galaxies, the predicted number of contaminated sources is N$_{p}$$\sim$0.3.
Thus, even under the most pessimistic assumptions no more than $\sim$ 
2/8 sources are contaminated.  

We obtained Hectospec (Fabricant et al. 2005) and FAST (Fabricant et al. 1998) spectra of these 8 sources
on the 6.5m MMT and 1.5m Tillinghast telescope at F. L. Whipple Observatory  
during September-November 2006.  
For the FAST sources, we took $\sim$ 10 minute exposures using
a 300 g mm$^{-1}$ grating blazed at 4750 $\AA$ and
a 3$\arcsec$ slit.  These spectra cover 3800--7500 $\AA$ at 6 $\AA$ resolution.  
For each Hectospec source (the six faintest), we took three, 10-minute exposures 
using the 270 g mm$^{-1}$ grating.  This configuration yields spectra at 
4000-9000$\AA$ with 3$\AA$ resolution.  The data were 
processed using standard FAST and Hectospec reduction pipelines (e.g. Fabricant et al. 2005).
%We wavelength-calibrate the spectra
%in NOAO IRAF\footnote{IRAF is distributed by the National Optical
%Astronomy Observatory, which is operated by the Association of
%Universities for Research in Astronomy, Inc. under contract to the
%National Science Foundation.}.
%After trimming the CCD frames, we correct for the bias level, flat-field each frame,
%apply an illumination correction, and derive a full wavelength
%solution from calibration lamps acquired immediately after each
%exposure.  The wavelength solution for each frame has a probable
%error of $\pm$0.5 $\dot{A}$ or better.  To construct final 1-D spectra,
%we extract object and sky spectra using the optimal extraction
%algorithm APEXTRACT within IRAF.  
The resulting spectra typically
have moderate signal-to-noise, S/N $\gtrsim$ 30 per pixel (Figure 1a). 
  
To measure spectral types, we derived
spectral indices of $H_{\alpha}$, $H_{\beta}$, $H_{\gamma}$, 
$H_{\delta}$, the G band (4305 $\dot{A}$), and Mg I (5175 $\dot{A}$) as in 
O'Connell (1973).  We derived piecewise linear relationships 
 between spectral indices and spectral type from the Jacoby et al. (1984) standards,  
similar to the method of Hernandez et al. (2004).  All sources have spectral 
types consistent with those expected for 13 Myr h \& $\chi$ Persei sources reddened by E(B-V)$\sim$ 
0.52 (Bragg \& Kenyon 2005) at a distance of 2.4 kpc. None are foreground M stars or background O/B stars.
No sources have a $H_{\alpha}$ equivalent width $\ge$10$\dot{A}$ indicative of 
gas accretion.  As an additional check on our spectral types, we fit the SED 
from U through $K_{s}$ to photospheric models in $\lambda$ vs. $\lambda$$F_{\lambda}$ space.   
Spectral types derived from spectroscopy and SED fitting agree to within 2-3 subclasses. 
To model their intrinsic SEDs through [8], the sources were dereddened in all 
bands.  We use Bessell \& Brett (1988) and Keller et al. (2001), respectively, to 
calibrate the V-I and U-B reddening from E(B-V), and
the IR extinction laws from Indebetouw (2005) to deredden the 2MASS and IRAC bands.  

Finally, we acquired MIPS 24 $\mu m$ photometry 
with an integration time of 80 seconds/pixel. 
The frames were processed using the MIPS Data Analysis Tool;
PSF fitting in the IRAF/DAOPHOT package was used to
obtain photometry adopting 7.3 Jy for the zero-point magnitude (Gordon et al. 2005).
We derive the MIPS 5$\sigma$ upper limits from the histogram of 
the number counts as a function of magnitude.  The MIPS data are complete to $\sim$ 10.5
and yield a 5$\sigma$ detection for only one source, source 5, 
with [24]$\sim$9.9$\pm$0.05.  
We estimate the [24] reddening from Mathis (1990): $A_{[24]}$$\approx$0.025.  
Therefore, our one 5$\sigma$ detection at [24] has $K_{s}$-[24]$\sim$4.4.  
%Table 1 shows a list of the sources to be analyzed.
\section{Modeling of Disk Emission}
%\subsection{Disk Models}
Figure 2 shows the source SEDs (filled diamonds).  To compare them with stellar 
photospheres, we used optical/IR colors for each spectral type 
from Kenyon \& Hartmann (1995; KH95) and converted from colors to
$\lambda$$F_{\lambda}$ in the Johnson-Cousins system. 
The zero-point flux from the models was scaled to match the  
dereddened V through J fluxes for each source (Allen 2001; Cohen et al. 2003).  
All sources have photospheric fluxes through [3.6]; one source (4) clearly has 
IR excess at [4.5].  All sources with [5.8] and [8] measurements show clear evidence 
for circumstellar emission.  The SEDs are 
similar to some 'anemic' disks around A-F stars in IC 348 (Lada et al. 2006).

To characterize the emission, we calculate the 
dust blackbody temperatures that match the observed [8] flux but do not overproduce
the flux at other wavelengths.
The range of dust blackbody temperatures (typically 300-400 K) for each source
imply terrestrial zone emission.  Derived disk luminosities are L$_{d}$/L$_{\star}$
$\sim$ 10$^{-4}$-10$^{-3}$, comparable to debris disks (R05).  

To investigate the disk emission in more detail, we constructed 
three disk models spanning a 
range of evolutionary states for disks surrounding $\sim$ 10 Myr old stars.  
The optically thick, non accreting, 
flat disk model (KH87) assumes that the disk 
passively reradiates photons from the star and extends 
from the magnetospheric truncation radius to $\gtrsim$ 2000 R$_{\star}$.
In a second disk model, an 'inner hole' model, all disk material $\le 50$ 
$R_{\star}$ ($\sim$ 1 AU for R=2 R$_{\odot}$, this was the best fit) 
from the star is removed from the optically thick, flat disk model.
This model is similar to the morphology expected for 'transition' T Tauri disks (e.g KH95).

Finally, we model the IR emission from an optically-thin debris disk.  
In a debris disk, modest IR excesses result from small grains produced by destructive collisions 
between $\sim 1-10$ km planetesimals during the oligarchic growth stage of planet 
formation (see KB04; Kenyon \& Bromley 2005).  As in the KB04 calculations, the starting  radial surface 
density ($\Sigma$) profile follows a minimum mass solar nebula, $\Sigma \propto r^{-3/2}$
 (Hayashi 1981).  Planetesimal collisions are modeled in the terrestrial zone, 
$\sim$ 1.5-7.5 AU away from a $2 M_{\odot}$ primary star.  Disk emission is tracked for 
$\sim 100$ Myr from the start of collisions.  The resulting SED is computed and 2MASS and 
IRAC colors are determined.

Figure 2 compares predictions from the three disk models to the data (Figure 2).
Optically-thin, debris disk models (dash-three dot line) 
consistently match the [5.8] and [8] IRAC fluxes.  While the debris 
disk models can overpredict the flux at [3.6] and [4.5] by 
$\sim$ 25$\%$, the predictions at [5.8] and [8] typically fall within 
20$\%$ of observed values.  In some cases (sources 2-4, 8) the agreement is excellent.
Standard optically-thick disk models (long dashes, source 1 only) 
and the best-fit optically-thick disk + inner hole models (dashes) fail to match the SEDs. 
 The optically-thick disk models overestimate the flux by $\ge$ a factor of 10 
beyond [3.6].  The inner hole model matches the observed SEDs through [3.6] but 
typically overpredicts the [5.8] ([8]) fluxes by factors of 1.5 (3).  
Furthermore, the optically-thick flat disk and inner hole models consistently 
predict a 24$\mu m$ flux $\sim$ 10-20$\times$ greater (or $\sim$ 3 magnitudes redder)
 than the observed flux (one source) or the 5$\sigma$ upper limits (5 sources).  
The debris disk model, which predicts $K_{s}$-[24]$\sim$3, 
is consistent with the 24 $\mu m$ upper limits of the sources 
not detected with MIPS.  The debris disk prediction is $\sim$ 2.5 times fainter
($\sim$ 1 magnitude bluer) than the observed $K_{s}$-[24] color from source 5.
However, planet formation calculations at 30-150 AU predict significant 24 $\mu m$ 
emission ($K_{s}$-[24] $\approx$ 3.5-4) by $\sim$ 10-30 Myr around 
2 $M_{\odot}$ stars (Currie et al. in prep., Kenyon \& Bromley, in prep.), so
regions beyond the terrestrial zone may contribute some [24] emission.
More massive disks or those with flatter density profiles (e.g. $r^{-1}$) than assumed here 
may also yield slightly larger [24] excesses.
%\subsection{IRAC colors \& MIPS photometry/upper limits}

Figure 3 shows the results in $K_{s}$-[4.5]/$K_{s}$-[8] and $K_{s}$-[5.8]/$K_{s}$-[8] 
color-color diagrams.  In both figures, the IRAC colors of our sources (diamonds/triangles) lie halfway 
between the photospheric and inner hole models (left square and dotted line).  
At the peak of debris disk emission (solid line), the 
debris disk model also produces IR colors about halfway between the photospheric and inner hole models. 
The colors show good agreement with debris disk predictions in 
$K_{s}$-[5.8] and $K_{s}$-[8], typically within 0.2 mags, and 
fair agreement with the [4.5] data.  
Inner hole models for earlier (later) spectral types predict $K_{s}$-[8] 
$\sim$ 1.75 (3), or $\sim$ 1-2 magnitudes too red.   
Many of the sources are, at best, marginally consistent
 with any star+background galaxy colors (dot-dashed enclosed region):
3/7 (4/7) are consistent with contamination in the $K_{s}$-[4.5]/$K_{s}$-[8] 
($K_{s}$-[5.8]/$K_{s}$-[8]) diagrams.  Only 2/8 are consistent with 
contamination in both plots.  Even considering systematic 
errors/uncertainties (model uncertainties, photometric 
errors), the sources occupy a very small region of possible star+galaxy colors 
and are still only marginally consistent with contamination.

The optically thick disk+inner hole model fails 
to model the IRAC/MIPS data in all cases; the debris disk model is consistent with 
the IRAC data and more consistent with the MIPS detection/upper limits.
Thus, debris from planet formation is the most plausible explanation for 
near-to-mid IR emission from these sources.  
\section{Discussion}
Near-to-mid IR-excess emission for our sample originates from dust 
with T$\sim$300-400 K and L$_{d}$/L$_{\star}$$\sim$10$^{-4}$-10$^{-3}$.
Circumstellar debris produced from terrestrial planet 
formation is the most likely explanation for this emission.
Future observations with the Spitzer Space Telescope and the James Webb 
Space Telescope can provide stronger constraints on the SEDs of these 
terrestrial zone debris disk candidates and greatly expand the 
sample of $\sim$ 13 Myr-old candidates in h and $\chi$ Persei.
Current IRAC/MIPS data require that the disks have little dust out to distances 
probed by [4.5] and are optically thin out to 
[24].  For a 13 Myr, K0 (F1) star with a luminosity of $\sim$ 1.49 (7.4) $L_{\odot}$, 
 the disks are dust free out to $\sim$ 0.23 (0.51) AU and optically thin to
$\sim$ 6.6 (14.6) AU.  Deeper MIPS [24] observations of these sources may 
more conclusively constrain the disk SED to compare with debris disk and transition disk models.  
Mid-IR spectroscopy may also aid source identification from the $\sim$ 10-20$\mu m$ flux levels and 
and the strength of the 10 $\mu m$ silicate feature (see 
Sicilia-Aguilar et al. 2007).  
\acknowledgements
We thank Matt Ashby, Rob Gutermuth, and Anil Seth for useful discussions regarding galaxy contamination and 
the anonymous referee for a thorough review.  We acknowledge support from the 
NASA Astrophysics Theory Program grant NAG5-13278, TPF Grant NNG06GH25G and the Spitzer GO program (Proposal 20132).  
T. C. received funding from an SAO Predoctoral Fellowship; Z. B. 
received support from 
Hungarian OTKA Grants TS049872, T042509, and T049082.
This work was supported by contract 1255094, issued by JPL/Caltech to the University 
of Arizona.  
%This research has made use of the NASA/IPAC Infrared Science Archive, which is 
%operated by the Jet Propulsion Laboratory, California Institute of Technology, under contract
%with the National Aeronautics and Space Administration.

\clearpage

\begin{deluxetable}{llllllllllllllll}
\tabletypesize{\scriptsize}
\tablecolumns{16}
%\tablecolumns{13}
\tablecaption{Sources analyzed in this paper (Dereddened, A$_{V}$$\sim$1.62)}
\tablehead{{$\#$}&{$\alpha$}&{$\delta$}&{ST}&{V}&{U-B}&{B-V}&{V-I}&{J}&{H}&{$K_{s}$}&{[3.6]}&{[4.5]}&{[5.8]}&{[8]}&{[24]}}
%\tablehead{{RA}&{DEC}&{V}&{U-B}&{B-V}&{V-I}&{J}&{H}&{$K_{s}$}&{[3.6]}&{[4.5]}&{[5.8]}&{[8]}}
\startdata
1&2:17:34.7&56:47:25 &F1&14.15& -    & -   &   -  &  13.58&  13.51&  13.52&  13.38$\pm$0.01&  13.29$\pm$0.05&  12.97$\pm$0.05&  12.15$\pm$0.04&-\\
2&2:17:38.0&56:53:51 &A9&14.03& 0.09& 0.17&   0.06& 13.61&  13.55&  13.45&  -&  -&  12.55$\pm$0.06&  12.35$\pm$0.14&-\\
3&2:19:11.3&57:09:53 &F1&14.72& 0.46& 0.37&   0.34& 13.88&  13.64&  13.64&  13.31$\pm$0.03&  13.23$\pm$0.02&  13.23$\pm$0.11& 12.89$\pm$0.08&$>$10.5\\
 4&2:22:06.4&57:07:06 &F7&15.44&-&  -& 0.51&  14.62&  14.22&  14.28&  -&  13.77$\pm$0.19&  -&  13.38$\pm$0.12&$>$10.5\\
 5&2:19:07.7&57:14:05 &F9&15.77& 0.35& 0.63&   0.53&  14.65&  14.26&  14.29&  14.27$\pm$0.06&  14.19$\pm$0.09&  13.83$\pm$0.10&  13.03$\pm$0.09&9.9\\
6&2:17:59.3&56:53:45 &G9&16.40&-&  -& 0.45& 15.15&  14.60&  14.87&  14.71$\pm$0.07&  14.73$\pm$0.11&  14.27$\pm$0.13&  14.06$\pm$0.19&$>$10.5\\
 7&2:17:39.5&57:20:26 &K0&16.43&-&  -& 0.76& 14.85&  14.43&  14.69&  14.43$\pm$0.06&  14.50$\pm$0.05&  14.29$\pm$0.18&  13.68$\pm$0.14&$>$10.5\\
 8&2:17:46.0&57:10:18 &K1&16.72&-&  -& 0.78& 15.12&  14.54&  14.43&  14.59$\pm$0.09&  14.35$\pm$0.10&  14.03$\pm$0.11&  13.51$\pm$0.12&$>$10.5\\
\enddata
%\tablecomments{Spectral types for sources 3-8 were derived from Hectospec observations and verified with SED fitting from 
%U through $K_{s}$.  Sources 1-2 were outside the Hectospec coverage, so the spectral types shown are those most consistent 
%with the source SEDs.  While these spectral types are somewhat uncertain, we note that the agreement is excellent between 
%spectral types from Hectospec and those most consistent with SED fitting.}
\end{deluxetable}

\clearpage

\begin{figure}
\epsscale{0.4}
\plotone{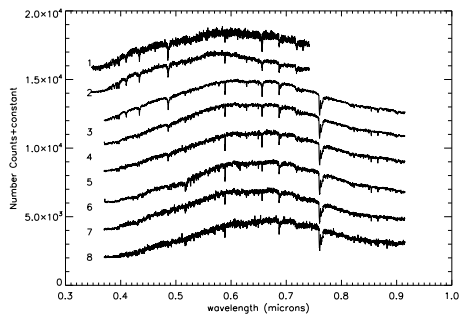}
%\plotone{Hspect.ps}
\caption{FAST and Hectospec spectra for our sample, with spectral types ranging from F1/A9 (top) to K0/K1 (bottom).  
The bottom source (8) has a 2000 count offset; spectra near the edges approach zero counts.}
\end{figure}
\begin{figure}
\epsscale{0.5}
%\plottwo{f2a.eps}{f2b.eps}
\plotone{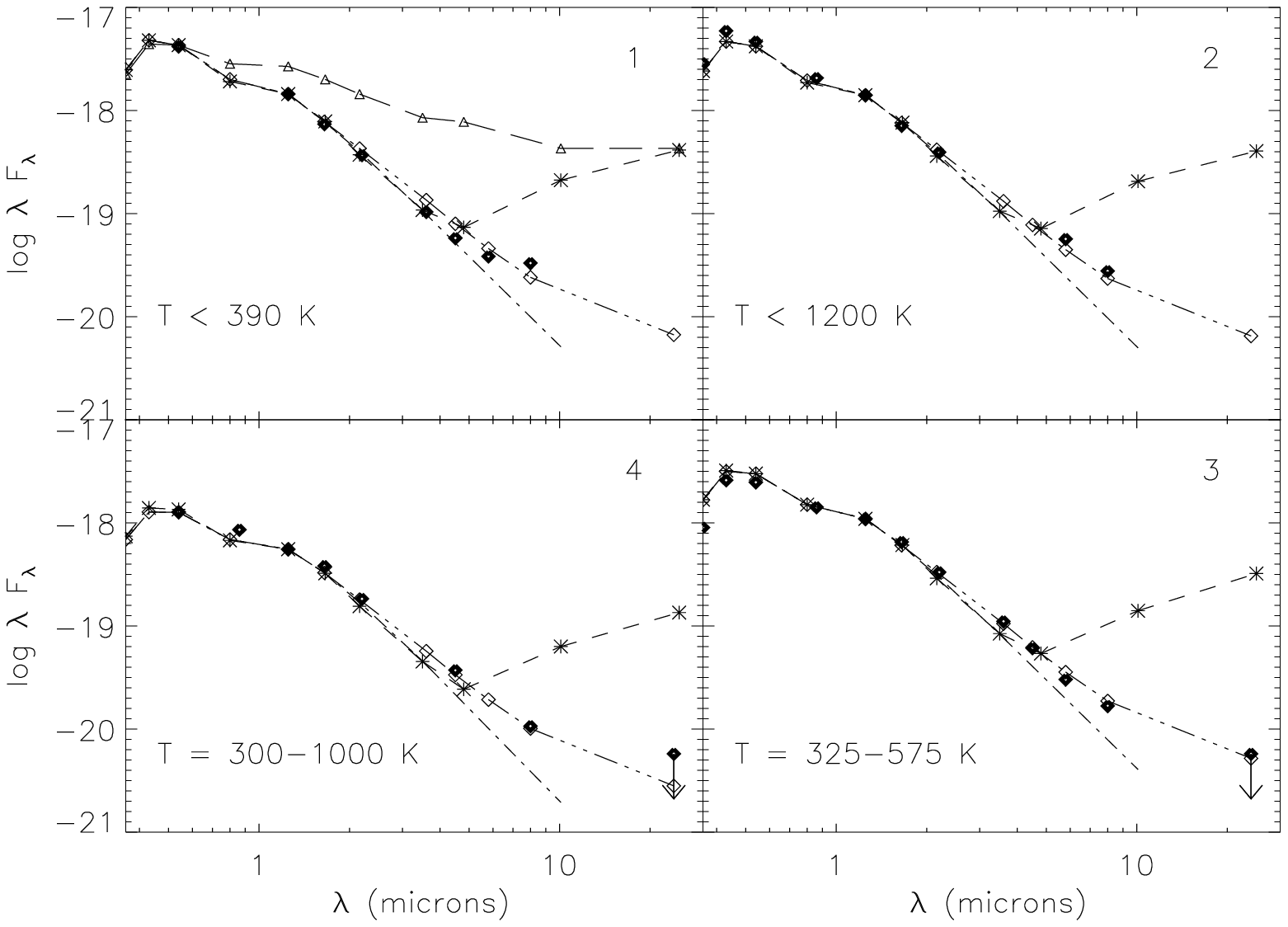}
\plotone{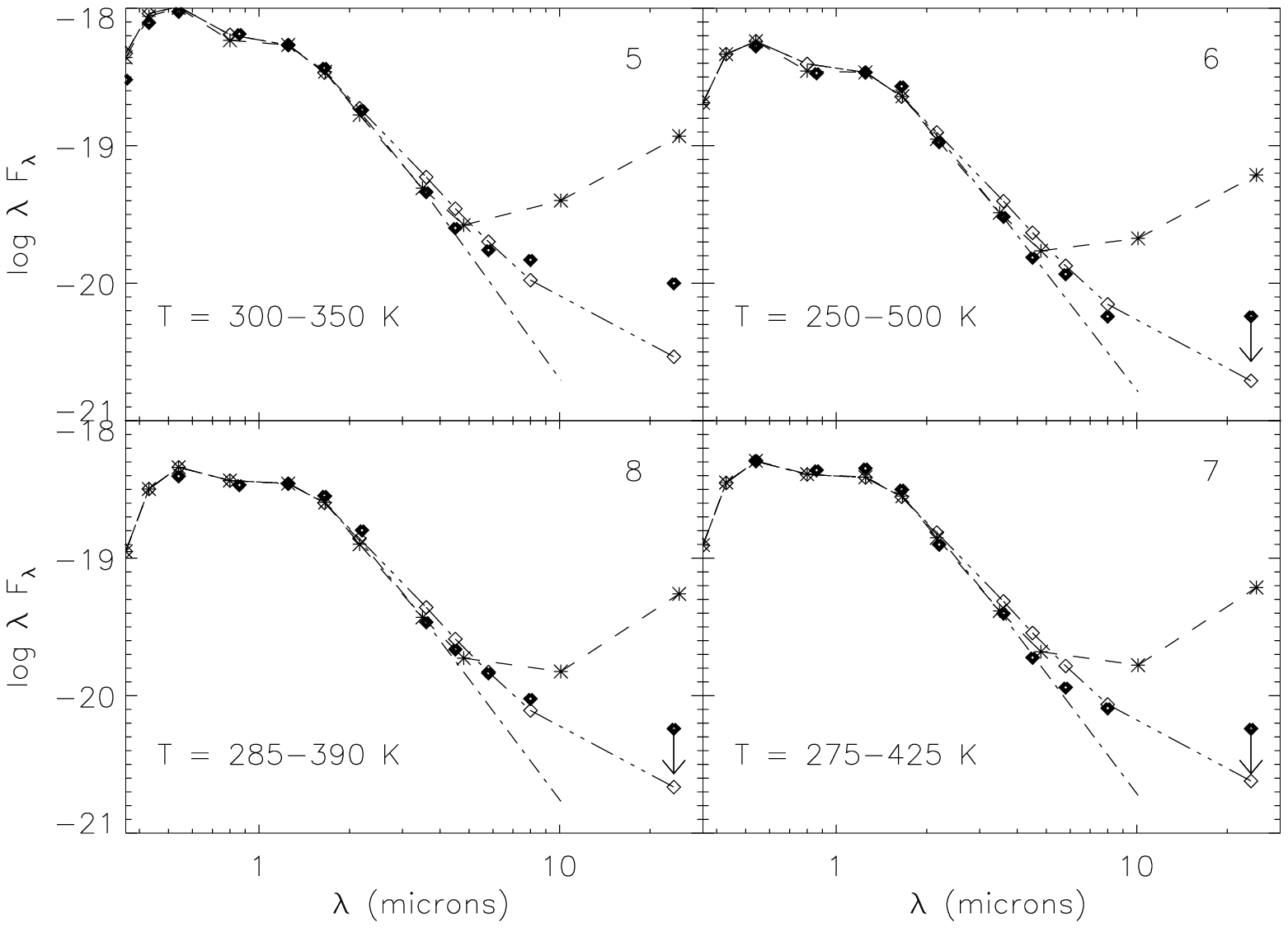}
\caption{Spectral energy distributions of the sources.  Filled diamonds denote 
detections; down arrows identify MIPS 5$\sigma$ upper limits.
The SED fits from V through [3.6] match photospheric predictions (dot-dashed line). 
Sources have clear, strong IR excess at [5.8] and [8].  
The dust blackbody temperatures indicate terrestrial zone emission.
Both standard optically-thick flat disk models (shown in the first plot only, large dashed line/open triangles) and 
optically-thick models with inner holes (dashed line/asterisks) overpredict the flux at [8].  Debris disk models (three dot-dashed line/open diamonds) appear to be 
consistent with the data.}
\end{figure}
%\clearpage
\begin{figure}
\epsscale{0.5}
\plotone{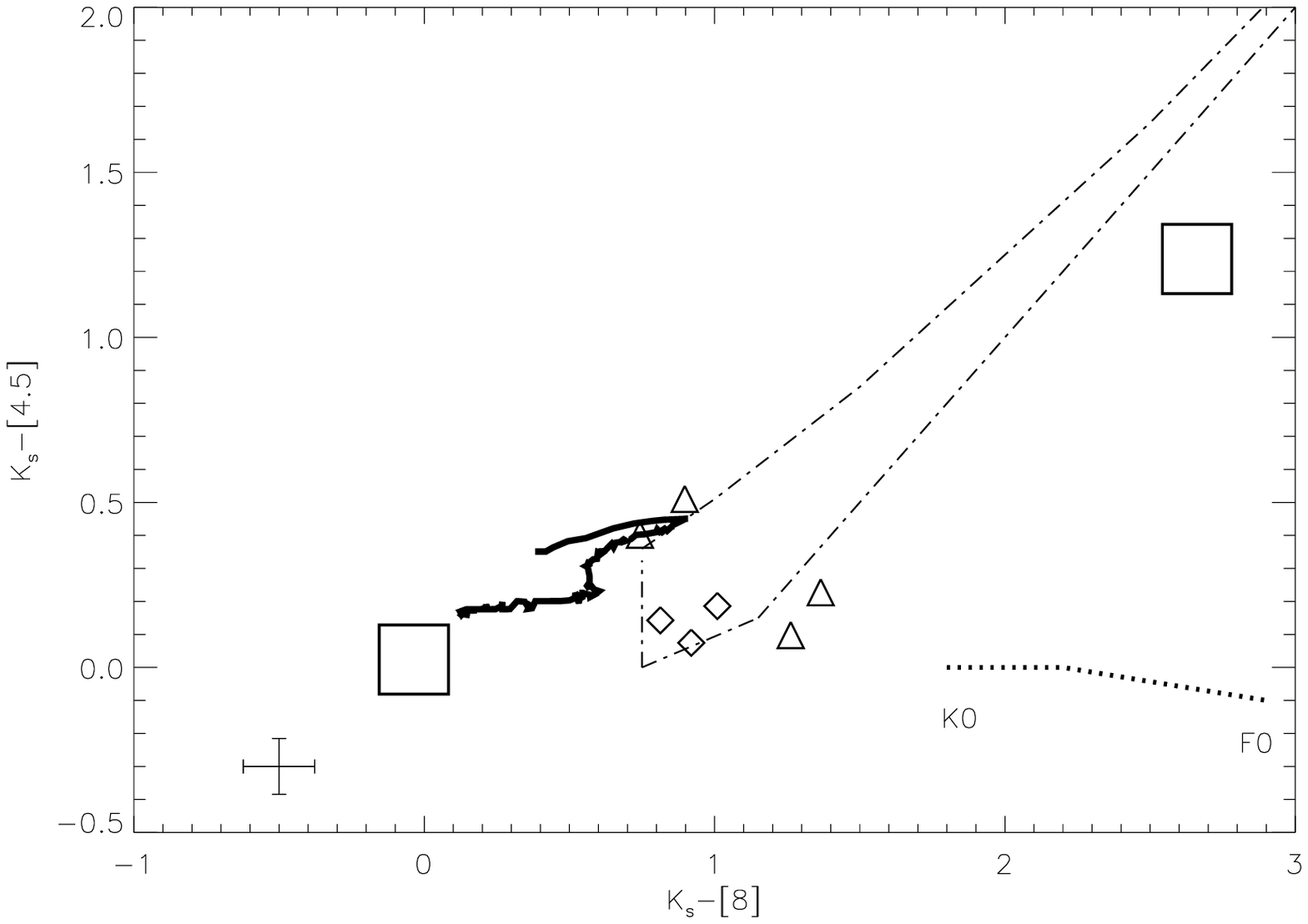}
\plotone{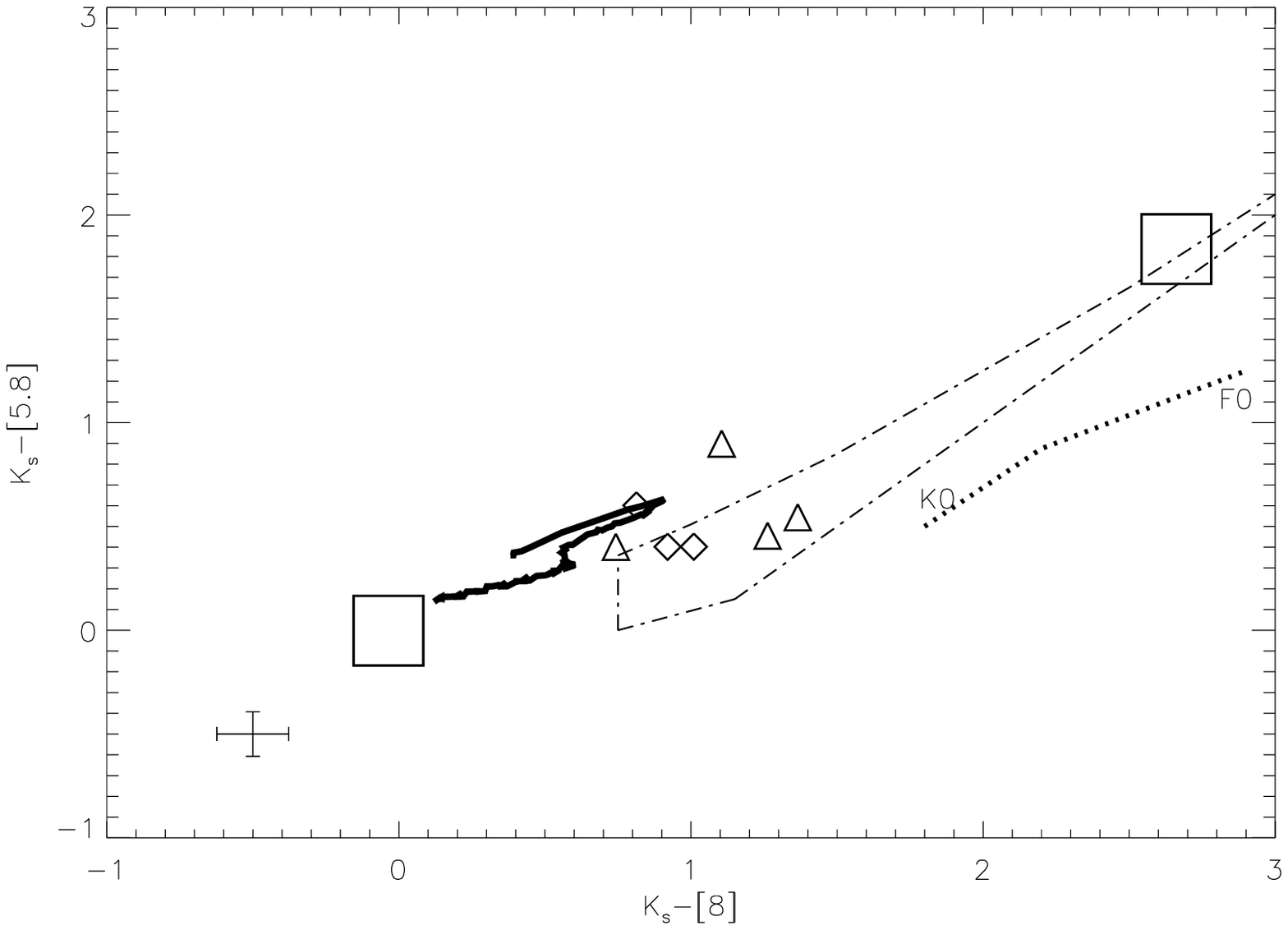}
\caption{$K_{s}$-[IRAC] colors for sources --triangles for sources earlier than G2, diamonds for sources later-- 
and predictions for photospheric, and standard optically-thick disk 
models (left and top-right squares), and a range of optically-thick disk + inner hole models (K0 to F0).  
The median errors in IRAC are shown in the lower left-hand corner.
The photospheric and optically-thick disk colors are for a spectral type of G2 and vary little with spectral type.
Overplotted as a thick solid line are the expected colors for emission produced by 
optically-thin debris, a byproduct of terrestrial planet formation.  The debris disk colors change due to 
the evolving dust production rate from planet formation.  The dot-dashed region encloses possible 
colors for photospheric sources+background galaxy contamination; PAH galaxies have [5.8]-[8]=1-3 and any 
star+galaxy composite 'source' with [4.5,5.8]-[8]$\gtrsim$1 would have been flagged as a contaminated source and 
removed from our sample.  
The debris disk model fits to within $\sim$ 0.3 mags in all cases; only 2/8 sources are consistent 
with star+PAH galaxy-contaminated colors.}
\end{figure}

\end{document}